\documentclass[review]{elsarticle}

\usepackage{lineno,hyperref}
\biboptions{sort&compress}
\modulolinenumbers[5]

\journal{Journal of \LaTeX\ Templates}

\usepackage{amsmath}

\begin{document}

\begin{frontmatter}

\title{CP sensitive observables of a hypothetical heavy spin-0 particle with $\gamma\gamma-$interactions dominant.} 
\author{N.~Belyaev}
\address{National Research Nuclear University MEPhI (Moscow Engineering Physics Institute), Kashirskoe highway 31, Moscow, 115409, Russia.}
\author{R.~Konoplich\fnref{myfootnote1}}
\address{Department of Physics, New York University, 4 Washington Place, New York, NY 10003, USA.}
\author{K.~Prokofiev}
\address{Department of Physics and Institute for Advanced Study, Hong Kong University of Science and Technology, Clear Water Bay, Kowloon, Hong Kong.}

\fntext[myfootnote1]{Also at Physics Department, Manhattan College, 4513 Manhattan College Parkway, Riverdale, New York, NY 10471, USA.}




\begin{abstract}
We study observables sensitive to tensor structure of interactions of a hypothetical heavy spin-0 particle.
It is assumed that the interactions of this particle are primary with photons; 
interactions with vector bosons $gg$, $WW$, $ZZ$, and quarks $t\bar{t}$  are 
suppressed. The above assumptions favor 
the production of this hypothetical particle through the vector boson fusion mechanism structurally 
dominated by $\gamma \gamma$ and $\gamma Z$ interactions. This particle will be produced
in association with two light quarks. It is shown that the difference in azimuthal angle between the 
tagging jets provides an observable sensitive to the CP properties of this hypothetical particle.

\end{abstract}

\begin{keyword}
 BSM \sep Heavy particle \sep  Tensor structure  \sep Extended Higgs sector \sep photon \sep jets \sep VBF \sep CP
\end{keyword}

\end{frontmatter}


\section{Introduction}
After the discovery of the Higgs boson by the ATLAS and CMS collaborations at the LHC \cite{Atlas_H, CMS_H}, 
the particle content of Standard Model (SM) is complete. Searches for physics beyond the 
Standard Model (BSM) are currently on-going. Several anomalies observed in semi-leptonic decays of B mesons 
\cite{Lees, Huschle, Aaij} could be better understood with the existence of a heavy resonance decaying
predominantly to photons \cite{Murphy}. Many BSM theories 
predict the existence of heavy neutral  resonances,  which could be produced in $pp$ collisions.
Several flavors of Two Higgs doublet (2HDM), composite Higgs, singlet Higgs and other models can accommodate  such  particles primarily being detected through their decays into photon pairs. 
Such a resonance could have  its couplings  to $t\bar{t}$ and $ZZ$ significantly
suppressed compared to $\gamma \gamma$.  The couplings to a $W^+W^-$
pair may be further suppressed or absent.  In the proposed models, the new particle may have CP-even, 
CP-odd or CP-mixed parity.  These properties lead to important consequences related 
to  the possible production  mechanism of such a hypothetical resonance.  The investigation of 
production mechanism  may thus help to reveal the  exact nature of such a particle.

In this paper we present a study of the properties of a new heavy neutral particle which could be discovered in high 
energy $pp$ collisions. We concentrate on the class of models where the di-photon  decays of such a particle will 
be the first decay mode to be observed experimentally. We further assume that this new 
hypothetical particle has spin $0$.  In case of observation of a heavy di-photon resonance, its spin-0
nature can be quickly investigated by  studying the distribution of its production angle $\cos \theta ^{\star}$, as suggested 
in \cite{Gao:2010qx}. In this paper we concentrate on the spin-0 scenario, investigating   CP-even, CP-odd and CP-mixed 
assumptions of parity by studying jet distributions from the production vertex of this resonance. 
Based on this model, we outline the experimental 
observables to be investigated to understand its exact nature. 

This article is organized as follows. Section~\ref{sec:models} provides a brief review of most relevant 
physics models capable of accommodating such a resonance. In Section~\ref{sec:vbf}
the corresponding modifications to the Vector Boson Fusion (VBF) production mechanism 
are discussed. In Section~\ref{sec:fs}
the characteristic jet distributions of such production are presented. The prospects for 
study is the exact nature of such a particle using these distributions are presented in Section~\ref{sec:summary}.

\section{\label{sec:models}Physics models}

Production in $pp$ collisions of a heavy neutral spin-0 resonance (we designate it as $S_{0}$) primarily detectable through its decay into 
pairs of photons corresponds to a relatively large class of theoretical models. Such resonances are 
predicted in certain flavors of two Higgs doublet (2HDM), composite Higgs models 
\cite{Lee:1973iz, Donoghue:1978cj, Dugan:1984hq, Cheng:1987rs, Agashe:2004rs, Martinez:2008hu, 
Branco:2011iw, Eberhardt:2013uba} and scalar singlet models
where $S_0$ couples to new vector-like quarks via a Yukawa coupling (see \cite{Falkow} and references therein). 
The common  property of the listed models is the presence of such a neutral spin-0 resonance with mass much 
larger than the SM Higgs boson mass. The production cross section of such a resonance would lie in the broad 
range of $30-6000$ fb, depending on 
the model and the resonance mass \cite{Heinemeyer:2013tqa}.

\begin{figure*}[h]
\begin{minipage}[h]{1\linewidth}
\centering
\center{\includegraphics[width=0.6\linewidth]{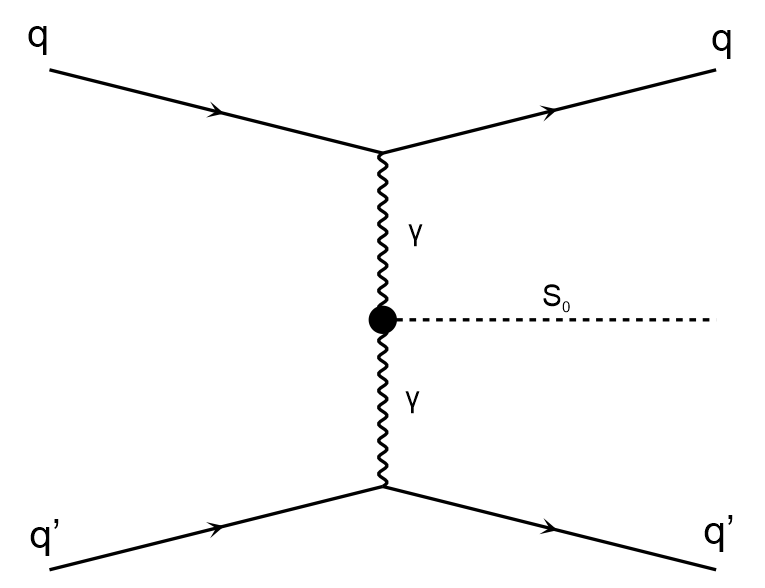}} 
\end{minipage}
\caption{Photon fusion feynman diagram. The $S_{0}$ particle can then decay through $\gamma\gamma$ and $Z\gamma$ channels.}
\label{fig:observables_45}
\end{figure*}.

In the listed models, $S_0$ couples relatively strongly to top quarks and photons. The dominant decay mode of $S_0$ is
thus  usually $t\bar{t}$, followed by the $\gamma\gamma$ mode. The third largest decay branching ratio is considered 
to be to  $Z\gamma$.  The decay branching ratio to $ZZ$ pairs in models with an extended Higgs sector  
can be $5-1000$ times smaller than the 
one corresponding to the  $\gamma\gamma$ decay \cite{Kanemura:2015bli, Diaz-Cruz:2014aga}.  
In the simplest formulation, decays to $W^+W^-$ can be further 
suppressed or absent. Observation of the $pp\to S\to t \bar{t}$ process appears to be difficult due to the large QCD background and to the complexity of
 $ t \bar{t}$ final state reconstruction. The most probable channel in which to observe such resonance in hadron collider experiments 
is thus $pp\to S_0\to \gamma \gamma$. The  $pp\to S_0\to Z\gamma$ decay observations should follow shortly after.

For some ranges of parameters of the models discussed above $t\bar{t}$ coupling can be also suppressed with respect to $\gamma\gamma$
coupling or even to be absent. The latter case can correspond to minimal singlet models with colorless vector-like fermions.   
The following study will be based on a minimal model with suppressed $ t \bar{t}$ coupling.
Given the potent coupling to photons, the dominant  $S_0$  production mode on hadron colliders in this model should be 
the photon-induced VBF production mechanism (see e.g. \cite{Falkow, Light, Strumia, Csaki1, Csaki2, Benbrik} and references therein) 
with small contributions 
from $Z\gamma$.  The kinematic properties of this photon-dominated VBF production mechanism are studied under the following physics model assumptions:
\begin{itemize}
 \item{ $S_0$ is a neutral s-channel resonance with mass about $1000$~GeV and spin $0$. Pure CP-even, CP-odd and CP-mixed parity 
 assumptions are considered. }
 \item{The effective coupling of $S_0$ to $\gamma\gamma$ is large compared to $Z\gamma$ and $ZZ$. The effective
 coupling to $ t \bar{t}$ and $W^+W^-$ is very small or absent. The $S_0$ VBF production is  dominated by 
 di-photon interactions with some contributions from $Z\gamma$  mechanisms. 
This process will be referred to as ``photon fusion" hereafter.}
\end{itemize}

\section{\label{sec:vbf} Photon fusion production}

Following the assumptions presented in the Section~\ref{sec:models},  the Effective Field Theory (EFT)
approach to describe the interactions of a spin-0 particle $S_0$ with vector bosons can be employed.
It is assumed that the masses of BSM particles which might contribute to loop-induced couplings of 
$S_0$ to SM particles are large compared to the $S_0$ mass. The corresponding interaction 
Lagrangian can be built by using a subset of $\gamma$ and $Z$-related dimension-5 operators. 
The resulting effective Lagrangian involves the following operators:
\begin{align}
 {\cal L}_0^V =\bigg\{
  & -\frac{1}{4}\big[\kappa_{S\gamma\gamma}
  A_{\mu\nu}A^{\mu\nu}
        +\kappa_{P\gamma\gamma}\,
 A_{\mu\nu}\widetilde A^{\mu\nu}
 \big] \nonumber \\
  &\mkern -40mu -\frac{1}{2}\big[\kappa_{SZ\gamma} \, 
 Z_{\mu\nu}A^{\mu\nu}
        +\kappa_{PZ\gamma}\,Z_{\mu\nu}\widetilde A^{\mu\nu} \big] \nonumber \\
  &\mkern -40mu -\frac{1}{\Lambda}
    \kappa_{S\partial\gamma} \, Z_{\nu}\partial_{\mu}A^{\mu\nu}     
 \bigg\} S_0\,,
 \label{eq:lagrange}
\end{align}
 where $\Lambda$ is the energy scale of new physics and the field strength tensors are defined as follows:
\begin{equation*}
V_{\mu\nu} =\partial_{\mu}V_{\nu}-\partial_{\nu}V_{\mu}\quad (V=A,Z).
\end{equation*}
The dual tensor $ \widetilde V_{\mu\nu}$ is defined as:
\begin{align}
 \widetilde V_{\mu\nu} =\frac{1}{2}\epsilon_{\mu\nu\rho\sigma}V^{\rho\sigma}.\,  \nonumber
\end{align}
The indices $S$ and $P$ represent respectively the  CP-even and CP-odd states of $S_0$.
The coupling constants $\kappa _{S\gamma\gamma}, \kappa _{SZ\gamma}, \kappa _{S\partial\gamma}, 
\kappa _{P\gamma\gamma}, \kappa _{PZ\gamma}$ govern  the strength of respective 
interactions with pairs of vector bosons. 
CP-violation is induced when both CP-odd and CP-even contributions are present simultaneously.
%

In the case of the gluon-fusion ($ggH$) process in the SM, the distribution 
of the azimuthal angle difference $\Delta \Phi = \Phi_2 - \Phi_1$ between the two tagging jets demonstrates 
a non-trivial structure sensitive to CP-properties of the Higgs boson 
\cite{Dolan:2014upa, Plehn:2001nj, Hankele:2006ma, Hagiwara:2009wt}. 
In the case of the electroweak SM VBF  process the distribution in $\Delta \Phi$ is relatively flat and thus is 
only barely sensitive to CP. This is the result of  the interference between $++$ and $--$ helicity 
states of vector bosons in gluon fusion Higgs boson production
and the dominance of the $00$ helicity state in the weak VBF case.    
Below we demonstrate that in models with suppressed $S_0ZZ$ and $S_0WW$ modes vector 
boson fusion processes could also reveal non-trivial azimuthal angle correlations. 

The tensor structure of $S_0VV$ vertices corresponding to the Lagrangian in Eq.~\ref{eq:lagrange} is given in
Table~\ref{tab:vertices}.
\begin{table}[htbp]
\centering
\begin{tabular}{lcc}
\hline
\hline
Interaction & CP & Tensor structure\\
\hline
$S_0\gamma\gamma, ~S_0Z\gamma$ & even & $g^{\mu\nu}(q_1q_2)-q_1^{\mu}q_2^{\nu}$\\
$S_0\gamma\gamma, ~S_0Z\gamma$ & odd & $\epsilon^{\mu\nu\alpha\beta}q_{1\alpha}q_{2\beta}$\\
$S_0\partial\gamma$ & even & $g^{\mu\nu}q_1^2-q_1^{\mu}q_1^{\nu}$\\
$HZZ ~SM$ & even & $ M_Z^2g^{\mu\nu} $\\
\hline
\hline
\end{tabular}
\caption{\label{tab:vertices} Tensor structure of vertices for $S_0$ interactions with vector bosons.
$q_1$ and $q_2$ are photon and vector boson four vectors, respectively.}
\end{table}
For comparison, the SM  $HZZ$ amplitude is also shown. 

Using the helicity amplitude technique \cite{Coradeschi:2012iu, Badger:2005jv} and following the approach developed 
in \cite{Hagiwara:2009wt} helicity amplitudes for $S_0VV$ vertices can be calculated. For a spin zero particle $S_0$
as a result of conservation of angular momentum only three amplitudes contribute to the VBF process:
$0 \to 00, 0 \to ++, 0 \to --$. These amplitudes are presented in Table~\ref{tab:amplitudes}.
\begin{table}[htbp]
\centering
\begin{tabular}{lccc}
\hline
\hline
Helicities & $ 00 $ & $ ++ $ & $ -- $ \\
\hline
$S_0\gamma\gamma, ~S_0Z\gamma$, even & $ (q_1^2q_2^2)^{1/2}  $ & $ -\frac{1}{2}M^2 $ & $ -\frac{1}{2}M^2 $ \\
$S_0\gamma\gamma, ~S_0Z\gamma$, odd & $ 0 $ & $ -\frac{i}{2}(M^2-4q_1^2q_2^2)^{1/2} $ & $ \frac{i}{2}(M^2-4q_1^2q_2^2)^{1/2} $ \\
$S_0\partial\gamma$, even & $ q_1^2\frac{M^2}{2(q_1^2q_2^2)^{1/2}} $ & $ q_1^2 $ & $ q_1^2 $ \\
$HZZ ~SM$, even & $M_Z^2\frac{M^2}{2(q_1^2q_2^2)^{1/2}} $ & $ -M_Z^2 $ & $ -M_Z^2 $\\
\hline
\hline
\end{tabular}
\caption{\label{tab:amplitudes} Helicity amplitudes for $S_0VV$ interactions. $M^2=M_{S_0}^2-q_1^2-q_2^2$}
\end{table}

In the VBF process absolute values of vector boson invariant masses are small in comparison with the mass of
$S_0$: $\sqrt{|q_i^2|} << M_{S_0}$. In this limit the 00-amplitude is dominant for the so-called contact 
term $k_{K\partial\gamma}$ and for the SM term. We cannot expect a non-trivial behavior of 
the azimuthal angle distribution for these cases. However in this limit the dominant amplitudes
for $S_0\gamma\gamma$ and $S_0Z\gamma$ vertices are the $++$ and $--$ amplitudes and their interference 
could lead to non-trivial azimuthal angle distributions in agreement with \cite{Hagiwara:2009wt}.          

In the case of CP-even and CP-odd $S_0$ particles the azimuthal angle distribution can be
presented in the form:
\begin{equation}
d\hat\sigma \sim A + Bcos(2\Delta\Phi),
\label{eq:dsig}
\end{equation}
where the coefficients A and B are obtained 
by combining helicity amplitudes for vector boson currents \cite{Hagiwara:2009wt} with the amplitudes given 
in Table~\ref{tab:amplitudes}. In the limit $\sqrt{|q_i^2|} << M_{S_0}$ these coefficients are presented 
in Table~\ref{tab:coef}, where $c_i=cos\Theta_i$ and $s_i=cos(\frac{\pi}{2}-\Theta_i)=sin\Theta_i$. Values $\Theta_i$ and $\frac{\pi}{2}-\Theta_i$ are correspond to the angles between 
directions of quarks and vector bosons in $q_{i}$ Breit frames.
\begin{table}[htbp]
\centering
\begin{tabular}{lcc}
\hline
\hline
Coefficients& $ A $ & $ B $ \\
\hline
$S_0\gamma\gamma, ~S_0Z\gamma$, even & $ q_1^2q_2^2M_{S_0}^4(1+c_1^2)(1+c_2^2)c_1^{-2}c_2^{-2} $ & $ q_1^2q_2^2M_{S_0}^4s_1^2s_2^2c_1^{-2}c_2^{-2} $ \\
$S_0\gamma\gamma, ~S_0Z\gamma$, odd & $ q_1^2q_2^2M_{S_0}^4(1+c_1^2)(1+c_2^2)c_1^{-2}c_2^{-2} $ & $ -q_1^2q_2^2M_{S_0}^4s_1^2s_2^2c_1^{-2}c_2^{-2} $  \\
$S_0\partial\gamma$, even & $ q_1^4M_{S_0}^4s_1^2s_2^2c_1^{-2}c_2^{-2} $ & $ \sim 0 $ \\
$HZZ ~SM$, even & $ M_Z^4M_{S_0}^4s_1^2s_2^2c_1^{-2}c_2^{-2} $ & $ \sim 0 $ \\
\hline
\hline
\end{tabular}
\caption{\label{tab:coef} Coefficients $A$ and $B$ of Eq.~\ref{eq:dsig}.}
\end{table}

According to this table $k_{S_0\partial\gamma}$ and the SM cases are dominated by the $A$ coefficient
in Eq.(\ref{eq:dsig}) resulting in a flat distribution over $\Delta\Phi$. For $S_0\gamma\gamma$ and $S_0Z\gamma$    
interactions the coefficients $A$ and $B$ are comparable leading to non-trivial correlation in $\Delta\Phi$.

\section{\label{sec:fs} Jet distributions}

To study the kinematic properties of the $S_0$ interactions with vector bosons, the corresponding Monte Carlo samples of 
500k events each were produced in leading order using MadGraph5 generator \cite{Alwall:2011uj}. The Higgs Characterisation model \cite{Artoisenet:2013puc}, implemented in MadGraph5, was 
used to study shapes of distributions in order to probe the possible effects of BSM physics.  
The parameters used in Monte Carlo production are listed in Table~\ref{tab:mc}.
\begin{table}[htbp]
\centering
\begin{tabular}{lc}
\hline
\hline
Parameter&Value\\
\hline
Mass of the resonance (GeV)& $m_{S_0} =1000$\\
Jet transverse momentum (GeV)& $p_T^{\rm jet}>30$\\
Jet pseudorapidity &$|\eta| < 4.0$ \\ 
\hline
\hline
\end{tabular}
\caption{\label{tab:mc} Parameters used in Monte Carlo production with the MadGraph5 generator.}
\end{table}

To estimate the effect of the presence of various contributions to the Lagrangian of Eq.(~\ref{eq:lagrange}) 
on the kinematics of the  
final state jets, the following final state observables are defined:
\begin{itemize}
 \item{The invariant mass of the final state jets: $m_{jj}$;}
 \item{The transverse momenta of $S_0$, the leading jet and the subleading jet, respectively: $p^{S_0}_{\rm T}$, 
$p^{\rm lead}_{\rm T}$, $p^{\rm sublead}_{\rm T}$;}
 \item{The pseudorapidities of the leading jet and subleading jet and the difference between jet 
pseudorapidities, respectively: $\eta_{\rm lead}$, $\eta_{\rm sublead}$, $|\eta| = |\eta_1 - \eta_2|$;}
 \item{The azimuthal angle difference between jets : 
$\Delta\Phi = \phi_1 - \phi_2$;}
 \item{The Zeppenfeld variable: $\left|\eta_{S_0} - \frac{\eta_1 + \eta_2}{2}\right|$.}
\end{itemize}

\begin{figure*}[htbp]
\begin{minipage}[h]{0.5\linewidth}
\center{\includegraphics[width=1\linewidth]{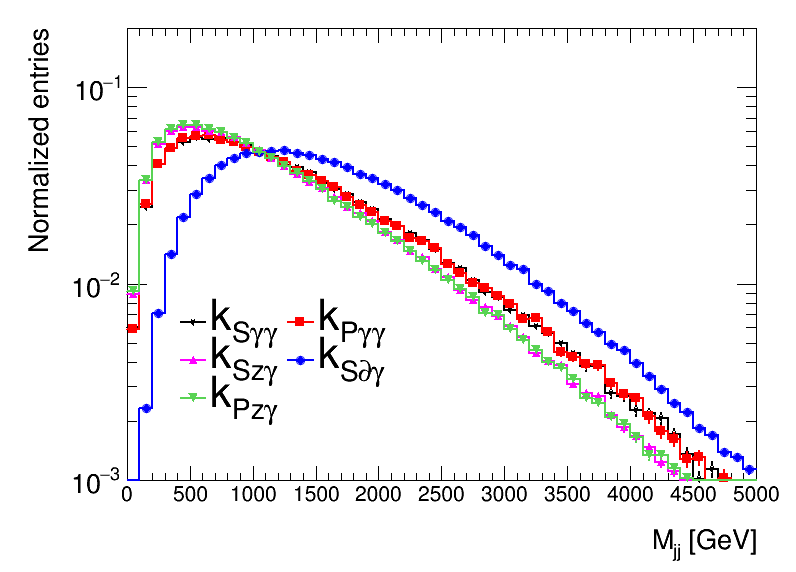} \\ (a)} 
\end{minipage}
\hfill
\begin{minipage}[h]{0.5\linewidth}
\center{\includegraphics[width=1\linewidth]{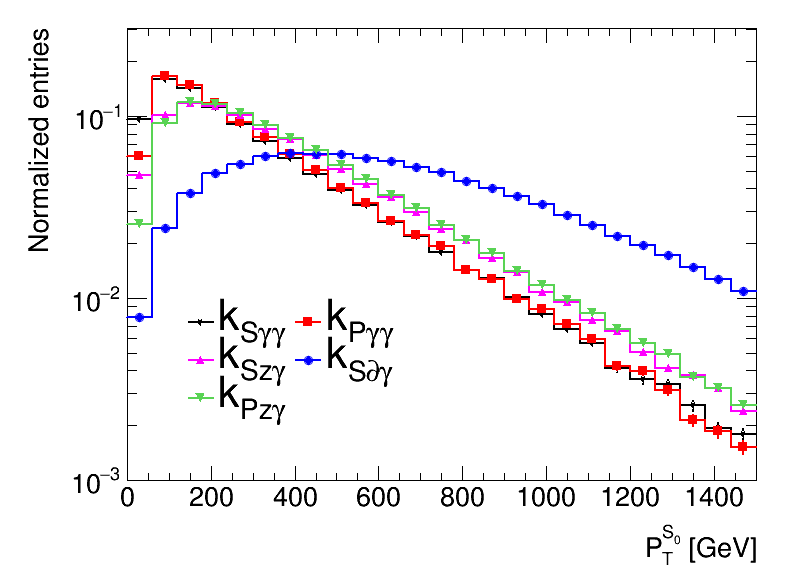} \\ (b)} 
\end{minipage}
\vfill
\begin{minipage}[h]{0.5\linewidth}
\center{\includegraphics[width=1\linewidth]{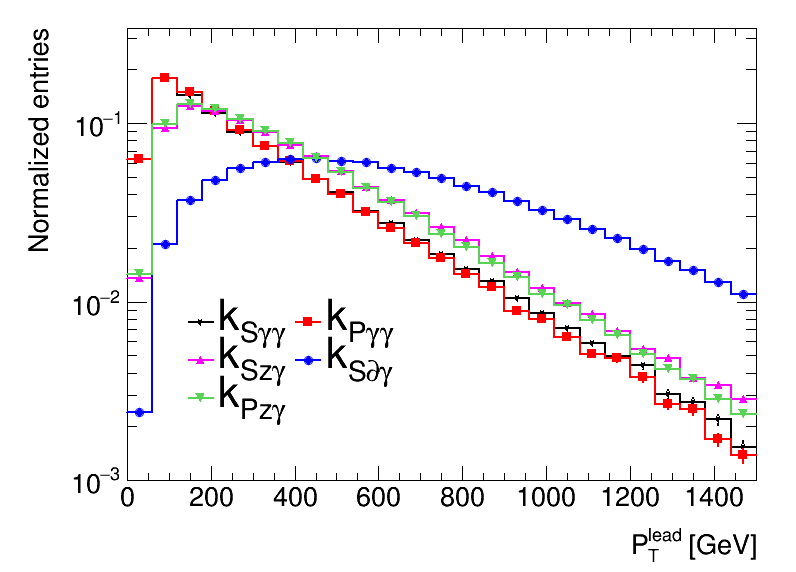} \\ (c)} 
\end{minipage}
\hfill
\begin{minipage}[h]{0.5\linewidth}
\center{\includegraphics[width=1\linewidth]{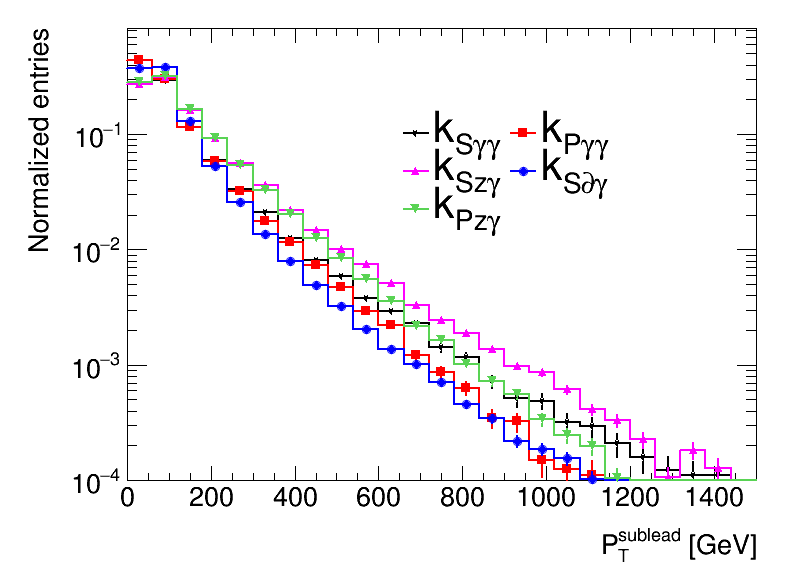} \\ (d)} 
\end{minipage}
\caption{Distributions of di-jet mass and the transverse momentum of tagging jets under various 
assumptions about the structure of the $S_0$ production mechanism.}
\label{fig:observables_1}
\end{figure*}

Fig.~\ref{fig:observables_1} shows the distributions of final state jet observables for various assumptions about the
structure of the $S_0$ production mechanism. In each case only one BSM operator is present at time; contributions 
from all other BSM operators are set to zero.

\begin{figure*}[htbp]
\begin{minipage}[h]{0.5\linewidth}
\center{\includegraphics[width=1\linewidth]{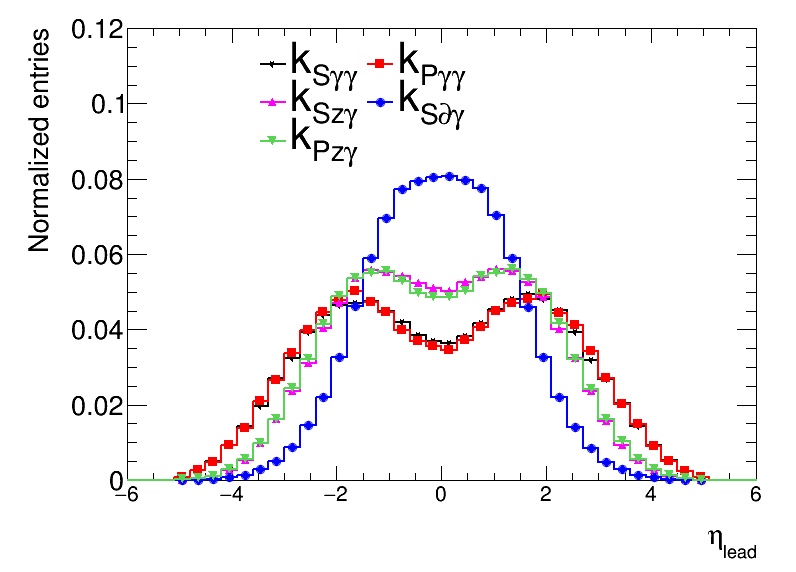} \\ (a)} 
\end{minipage}
\hfill
\begin{minipage}[h]{0.5\linewidth}
\center{\includegraphics[width=1\linewidth]{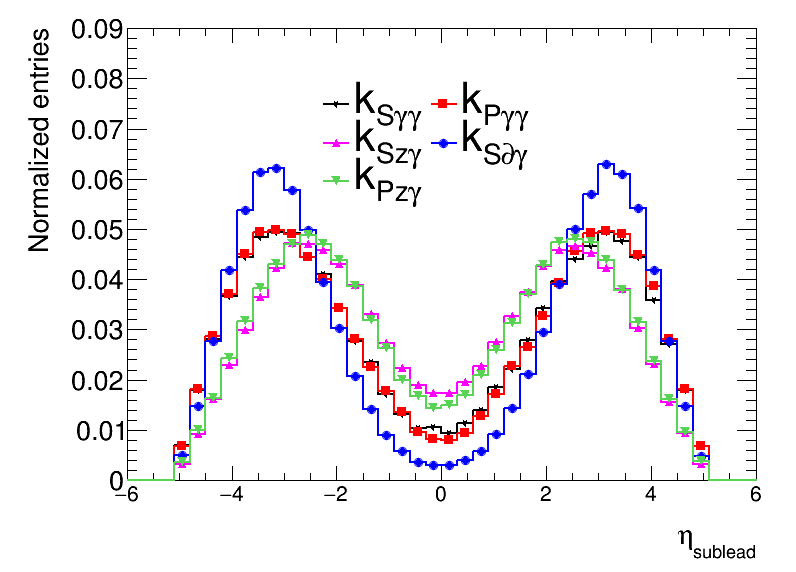} \\ (b)} 
\end{minipage}
\vfill
\begin{minipage}[h]{0.5\linewidth}
\center{\includegraphics[width=1\linewidth]{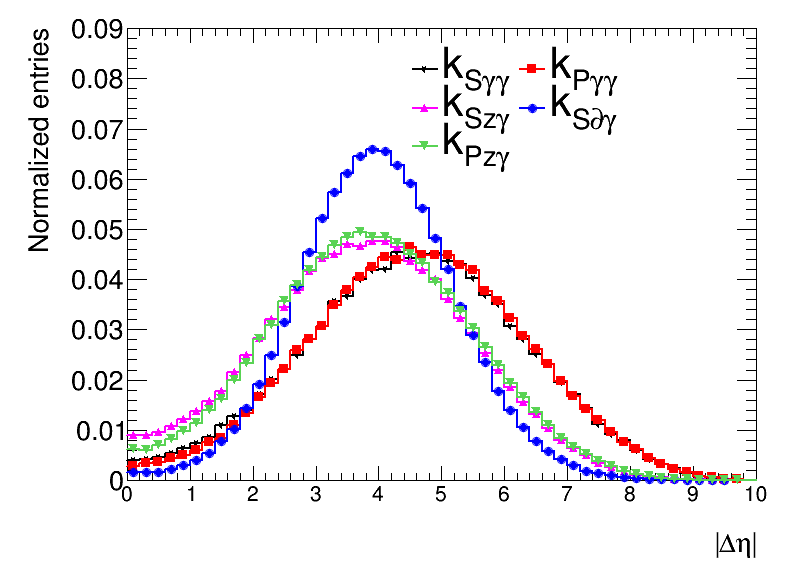} \\ (c)} 
\end{minipage}
\hfill
\begin{minipage}[h]{0.5\linewidth}
\center{\includegraphics[width=1\linewidth]{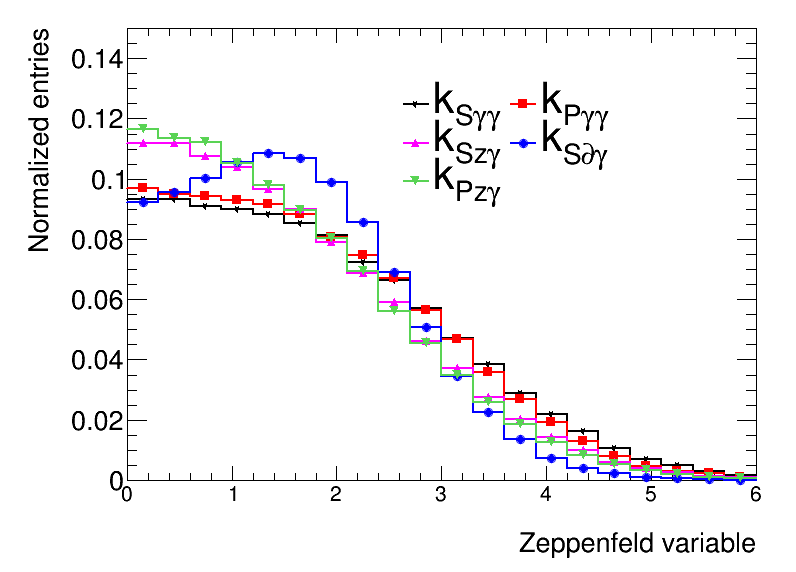} \\ (d)} 
\end{minipage}
\caption{Distributions of pseudorapidities of tagging jets under  various 
assumptions about the  structure of the $S_0$ production mechanism.}
\label{fig:observables_2}
\end{figure*}
These observables can provide some sensitivity to the presence of an operator 
corresponding to the coupling $k_{S\partial\gamma}$.
The presence of operators corresponding to $k_{S\gamma\gamma}$ and $k_{P\gamma\gamma}$ 
results in nearly indistinguishable distributions for all $P_T$ observables. 
A similar situation arises for the case when only operators corresponding to $k_{SZ\gamma}$ 
and $k_{PZ\gamma}$ couplings are present.
 
In Fig.~\ref{fig:observables_2} pseudorapidity distributions of jets are presented. These distributions 
demonstrate the prominent feature of VBF processes: the suppression of central jets and the appearance
of a central pseudorapidity gap. The only exception here is the leading jet distribution corresponding 
to the term $k_{S\partial\gamma}$. 
 

Note that a comparison of gluon fusion and photon fusion heavy resonance production mechanisms on the basis of
$P_T$ and $\eta$ distributions was performed in \cite{Csaki2}.

\begin{figure*}[htbp]
\begin{minipage}[h]{0.5\linewidth}
\center{\includegraphics[width=1\linewidth]{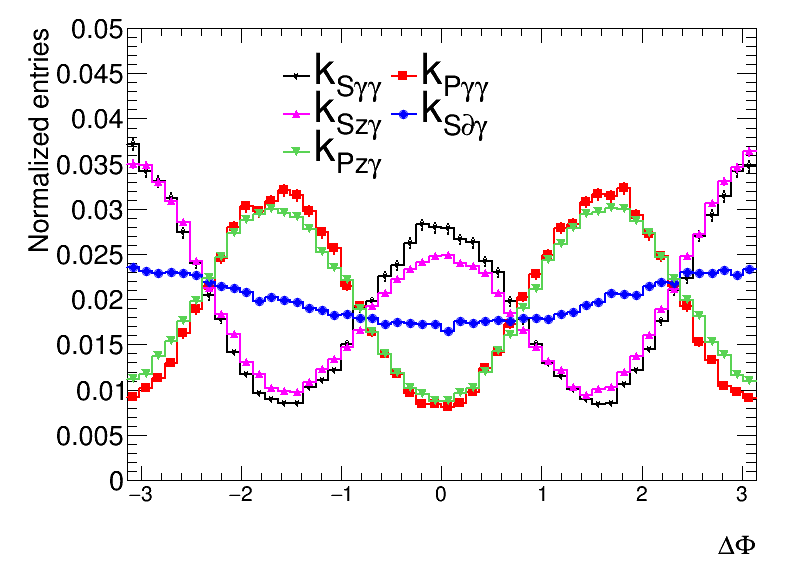} \\ (a)} 
\end{minipage}
\hfill
\begin{minipage}[h]{0.5\linewidth}
\center{\includegraphics[width=1\linewidth]{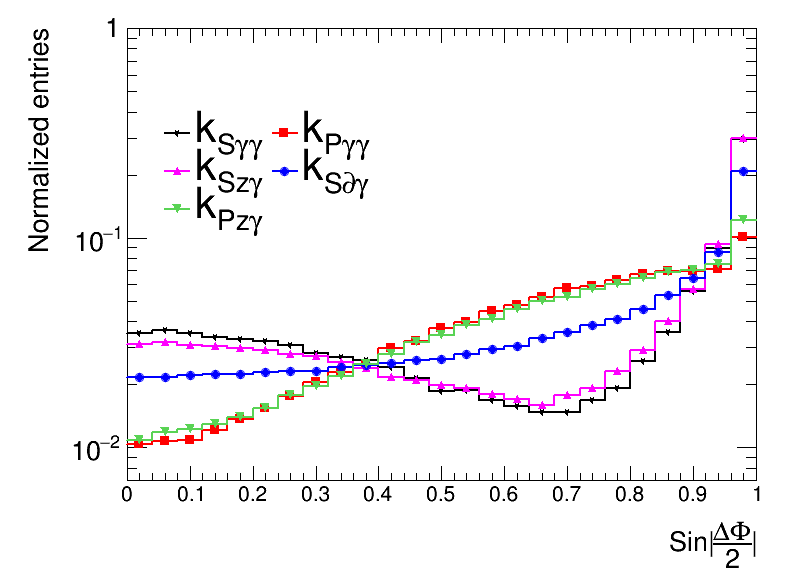} \\ (b)} 
\end{minipage}
\caption{The azimuthal angle difference distributions between the tagging jets.}
\label{fig:observables_3}
\end{figure*}

The observable sensitive to CP properties of a heavy resonance produced via photon fusion is 
the azimuthal angle difference between the tagging jets.
In Fig.~\ref{fig:observables_3} $\Delta\Phi$ and $\sin\left|\frac{\Delta\Phi}{2}\right|$ distributions
are presented for the CP-even, CP-odd and CP-mixed cases. 

A clear separation in shapes of the observables is visible for cases with different CP-parity.
The CP-mixing examples are produced by requiring the simultaneous presence of operators
corresponding to the $k_{S\gamma\gamma}$ and $k_{P\gamma\gamma}$ terms. 
The terms corresponding to the $k_{S\gamma\gamma}$   and   $k_{P\gamma\gamma}$ couplings contribute $2/3$ and  $1/3$ 
of the total cross-section, respectively.  In the mixed CP-case, an additional term proportional to  $\sin\left|\frac{\Delta\Phi}{2}\right|$ appears in Eq.~\ref{eq:dsig} leading to a shift  in the distribution of $\sin\left|\frac{\Delta\Phi}{\
2}\right|$  as shown in Fig.~\ref{fig:observables_3}(b). 
This is an important property of photon fusion that can be used to reveal the structure of $Z\gamma$ and $\gamma\gamma$ vertices 
in current and future collider experiments.
\clearpage
\section{\label{sec:summary} Conclusion}
In this paper we study CP sensitive observables and the tensor structure of interactions of a hypothetical heavy 
spin-0 particle $S_0$. We assume that the effective coupling of $S_0$ to $\gamma\gamma$ is large 
compared to $t\bar{t}, ZZ$ and $WW$. Particles with these properties appear in models with an extended Higgs
sector. 

We focus on photon fusion production of $S_0$ and study various jet distributions and correlations between 
tagging jets from the production vertex. In the SM, jets produced via VBF are weakly correlated resulting in
a flat distribution in the azimuth angle difference $\Delta\Phi$ between the jets and limited sensitivity to the CP properties 
of $S_0$. On the other hand, we demonstrate that in the case of photon fusion the higher dimension 
operators lead to non-trivial $\Delta\Phi$ dependence and sensitivity to CP properties. This result is 
independent of decay processes because for a spin-0 s-channel resonance the production and decay vertices are
decoupled. 

Production jet distributions in photon fusion are also sensitive to a specific type of higher dimension 
operators. In particular, we show that the contact term proportional to $k_{S\partial\gamma}$ in the Lagrangian 
of the model considered can be distinguished from other BSM terms by analyzing $P_T, \eta$ and $\Delta\Phi$
distributions.    

In this paper we consider a model with a negligible $S_0t\bar{t}$ interaction. This is not the case when the dominant
$S_0$ production is gluon fusion. In analogy with the SM, $\Delta\Phi$ distribution of jets produced in
gluon fusion would have CP sensitivity. If gluon fusion and photon fusion production processes for $S_0$ could be 
effectively separated then the analysis of $\Delta\Phi$ distributions for these two processes would reveal 
important information about the tensor structure of $S_0$ interactions.  

ACKNOWLEDGMENTS.
The work of R. Konoplich is partially supported by the US National Science Foundation under Grant No.PHY-1402964. 
The work of K. Prokofiev is partially supported by a grant from the Research Grant Council of the Hong Kong 
Special Administrative Region, China (Project Nos. CUHK4/CRF/13G). The work of N. Belyaev was performed within the framework of the Center for Fundamental Research and Particle Physics supported by the MEPhI Academic Excellence Project (contract № 02.a03.21.0005, 27.08.2013). 

\section*{}


\begin{thebibliography}{10}
\expandafter\ifx\csname url\endcsname\relax
  \def\url#1{\texttt{#1}}\fi
\expandafter\ifx\csname urlprefix\endcsname\relax\def\urlprefix{URL }\fi
\expandafter\ifx\csname href\endcsname\relax
  \def\href#1#2{#2} \def\path#1{#1}\fi

\bibitem{Atlas_H}
G.~{Aad et al.}, {Observation of a new particle in the search for the Standard
  Model Higgs boson with the ATLAS detector at the LHC}, Phys. Lett. B 716
  (2012) 1--29.
\newblock \href {http://dx.doi.org/10.1016/j.physletb.2012.08.020}
  {\path{doi:10.1016/j.physletb.2012.08.020}}.

\bibitem{CMS_H}
S.~{Chatrchyan et al.}, {Observation of a new boson at a mass of 125 GeV with
  the CMS experiment at the LHC}, Phys. Lett. B716 (2012) 30--61.
\newblock \href {http://dx.doi.org/10.1016/j.physletb.2012.08.021}
  {\path{doi:10.1016/j.physletb.2012.08.021}}.

\bibitem{Lees}
J.~P. {Lees and et al.}, {Measurement of an Excess of $\bar{B} \to
  D^{(*)}\tau^- \bar{\nu}_\tau$ Decays and Implications for Charged Higgs
  Bosons}, Phys. Rev. D 88~(7) (2013) 072012.
\newblock \href {http://dx.doi.org/10.1103/PhysRevD.88.072012}
  {\path{doi:10.1103/PhysRevD.88.072012}}.

\bibitem{Huschle}
M.~{Huschle et al.}, {Measurement of the branching ratio of $\bar{B} \to
  D^{(\ast)} \tau^- \bar{\nu}_\tau$ relative to $\bar{B} \to D^{(\ast)} \ell^-
  \bar{\nu}_\ell$ decays with hadronic tagging at Belle}, Phys. Rev. D 92~(7)
  (2015) 072014.
\newblock \href {http://dx.doi.org/10.1103/PhysRevD.92.072014}
  {\path{doi:10.1103/PhysRevD.92.072014}}.

\bibitem{Aaij}
R.~{Aaij et al.}, {Measurement of the ratio of branching fractions
  $\mathcal{B}(\bar{B}^0 \to
  D^{*+}\tau^{-}\bar{\nu}_{\tau})/\mathcal{B}(\bar{B}^0 \to
  D^{*+}\mu^{-}\bar{\nu}_{\mu})$}, Phys. Rev. Lett. 115~(11) (2015) 111803.
\newblock \href {http://dx.doi.org/10.1103/PhysRevLett.115.159901}
  {\path{doi:10.1103/PhysRevLett.115.159901}}.

\bibitem{Murphy}
C.~Murphy, {Vector Leptoquarks and the 750 GeV Diphoton Resonance at the LHC},
  Phys. Lett. B 757 (2016) 192--198.
\newblock \href {http://dx.doi.org/10.1016/j.physletb.2016.03.076}
  {\path{doi:10.1016/j.physletb.2016.03.076}}.

\bibitem{Gao:2010qx}
Y.~Gao, A.~Gritsan, Z.~{Guo et al.}, Spin determination of single-produced
  resonances at hadron colliders, Phys. Rev. D 81 (2010) 075022.
\newblock \href {http://dx.doi.org/10.1103/PhysRevD.81.075022}
  {\path{doi:10.1103/PhysRevD.81.075022}}.

\bibitem{Lee:1973iz}
T.~D. Lee, A theory of spontaneous T violation, Phys. Rev. D 8 (1973)
  1226--1239.
\newblock \href {http://dx.doi.org/10.1103/PhysRevD.8.1226}
  {\path{doi:10.1103/PhysRevD.8.1226}}.

\bibitem{Donoghue:1978cj}
J.~Donoghue and L.~Li, Properties of charged Higgs bosons, Phys. Rev. D 19 (1979)
  945.
\newblock \href {http://dx.doi.org/10.1103/PhysRevD.19.945}
  {\path{doi:10.1103/PhysRevD.19.945}}.

\bibitem{Dugan:1984hq}
M.~Dugan, H.~Georgi and D.~Kaplan, Anatomy of a Composite Higgs model, Nucl. Phys.
  B 254 (1985) 299--326.
\newblock \href {http://dx.doi.org/10.1016/0550-3213(85)90221-4}
  {\path{doi:10.1016/0550-3213(85)90221-4}}.

\bibitem{Cheng:1987rs}
T.~P. Cheng and M.~Sher, Mass matrix ansatz and flavor nonconservation in models
  with multiple Higgs doublets, Phys. Rev. D 35 (1987) 3484.
\newblock \href {http://dx.doi.org/10.1103/PhysRevD.35.3484}
  {\path{doi:10.1103/PhysRevD.35.3484}}.

\bibitem{Agashe:2004rs}
K.~Agashe, R.~Contino and A.~Pomarol, The Minimal composite Higgs model, Nucl.
  Phys. B 719 (2005) 165--187.
\newblock \href {http://dx.doi.org/10.1016/j.nuclphysb.2005.04.035}
  {\path{doi:10.1016/j.nuclphysb.2005.04.035}}.

\bibitem{Martinez:2008hu}
R.~Martinez, J.~Rodriguez and S.~Sanchez, Charged Higgs production at photon
  colliders in 2HDM-III, Braz. J. Phys. 38 (2008) 507--510.
\newblock \href {http://dx.doi.org/10.1590/S0103-97332008000400025}
  {\path{doi:10.1590/S0103-97332008000400025}}.

\bibitem{Branco:2011iw}
G.~Branco, P.~Ferreira, L.~{Lavoura et al.}, Theory and phenomenology of
  Two-Higgs-doublet models, Phys. Rept. 516 (2012) 1--102.
\newblock \href {http://dx.doi.org/10.1016/j.physrep.2012.02.002}
  {\path{doi:10.1016/j.physrep.2012.02.002}}.

\bibitem{Eberhardt:2013uba}
O.~Eberhardt, U.~Nierste and M.~Wiebusch, Status of the Two-Higgs-doublet model of
  type II, JHEP 07 (2013) 118.
\newblock \href {http://dx.doi.org/10.1007/JHEP07(2013)118}
  {\path{doi:10.1007/JHEP07(2013)118}}.

\bibitem{Falkow}
A.~Falkowski, O.~Slone and T.~Volansky, {Phenomenology of a 750 GeV Singlet}, JHEP
  02 (2016) 152.
\newblock \href {http://dx.doi.org/10.1007/JHEP02(2016)152}
  {\path{doi:10.1007/JHEP02(2016)152}}.

\bibitem{Heinemeyer:2013tqa}
{The LHC Higgs Cross Section Working Group}, Handbook of LHC Higgs cross
  sections: 3. Higgs properties (2013).
\newblock \href {http://dx.doi.org/10.5170/CERN-2013-004}
  {\path{doi:10.5170/CERN-2013-004}}.

\bibitem{Kanemura:2015bli}
S.~Kanemura, K.~Nishiwaki, H.~Okada and J.A.~Orduz-Ducuara, {LHC 750 GeV Diphoton excess in a
  radiative Seesaw model} (2015).
	\newblock \href {https://arxiv.org/abs/1512.09048}
  {\path{arXiv:1512.09048}}.

\bibitem{Diaz-Cruz:2014aga}
J.~Diaz-Cruz, C.~Honorato, J.~Orduz-Ducuara and M. A. Pérez, One-loop decays $a^0 \to
  zz, z\gamma, \gamma\gamma$ within the 2HDM and its search at the LHC, Phys.
  Rev. D 90~(9) (2014) 095019.
\newblock \href {http://dx.doi.org/10.1103/PhysRevD.90.095019}
  {\path{doi:10.1103/PhysRevD.90.095019}}.

\bibitem{Light}
S.~Fichet, G.~Gersdorff and C.~Royon, {Scattering light by light at 750 GeV at the
  LHC}, Phys. Rev. D 93~(7) (2016) 075031.
\newblock \href {http://dx.doi.org/10.1103/PhysRevD.93.075031}
  {\path{doi:10.1103/PhysRevD.93.075031}}.

\bibitem{Strumia}
R.~Franceschini, G.~Giudice, J.~{Kamenik et al.}, {What is the $\gamma \gamma$
  resonance at 750 GeV?}, JHEP 03 (2016) 144.
\newblock \href {http://dx.doi.org/10.1007/JHEP03(2016)144}
  {\path{doi:10.1007/JHEP03(2016)144}}.

\bibitem{Csaki1}
C.~Cs{\'a}ki, J.~Hubisz and J.~Terning, {Minimal model of a Diphoton resonance:
  Production without gluon couplings}, Phys. Rev. D 93~(3) (2016) 035002.
\newblock \href {http://dx.doi.org/10.1103/PhysRevD.93.035002}
  {\path{doi:10.1103/PhysRevD.93.035002}}.

\bibitem{Csaki2}
C.~Cs{\'a}ki, J.~Hubisz, S.~Lombardoand and J.~Terning, {Gluon versus photon
  production of a 750 GeV Diphoton resonance}, Phys. Rev. D 93~(9) (2016)
  095020.
\newblock \href {http://dx.doi.org/10.1103/PhysRevD.93.095020}
  {\path{doi:10.1103/PhysRevD.93.095020}}.

\bibitem{Benbrik}
R.~Benbrik, C.-H. Chen and T.~Nomura, {Higgs singlet boson as a Diphoton resonance
  in a vectorlike quark model}, Phys. Rev. D 93~(5) (2016) 055034.
\newblock \href {http://dx.doi.org/10.1103/PhysRevD.93.055034}
  {\path{doi:10.1103/PhysRevD.93.055034}}.

\bibitem{Dolan:2014upa}
M.~Dolan, P.~Harris, M.~Jankowiak and M.~Spannowsky, Constraining $CP$-violating Higgs
  sectors at the LHC using gluon fusion, Phys. Rev. D90 (2014) 073008.
\newblock \href {http://dx.doi.org/10.1103/PhysRevD.90.073008}
  {\path{doi:10.1103/PhysRevD.90.073008}}.

\bibitem{Plehn:2001nj}
T.~Plehn, D.~Rainwater and D.~Zeppenfeld, Determining the structure of Higgs
  couplings at the LHC, Phys. Rev. Lett. 88 (2002) 051801.
\newblock \href {http://dx.doi.org/10.1103/PhysRevLett.88.051801}
  {\path{doi:10.1103/PhysRevLett.88.051801}}.

\bibitem{Hankele:2006ma}
T. Figy, V.~Hankele, G.~Kl{\"a}mke, D.~Zeppenfeld, Anomalous Higgs boson couplings
  in vector boson fusion at the CERN LHC, Phys. Rev. D 74 (2006) 095001.
\newblock \href {http://dx.doi.org/10.1103/PhysRevD.74.095001}
  {\path{doi:10.1103/PhysRevD.74.095001}}.

\bibitem{Hagiwara:2009wt}
K.~Hagiwara, Q.~Li and K.~Mawatari, Jet angular correlation in vector-boson fusion
  processes at hadron colliders, JHEP 07 (2009) 101.
\newblock \href {http://dx.doi.org/10.1088/1126-6708/2009/07/101}
  {\path{doi:10.1088/1126-6708/2009/07/101}}.

\bibitem{Coradeschi:2012iu}
F.~Coradeschi and P.~Lodone, Selection rules for helicity amplitudes in massive
  gauge theories, Phys. Rev. D87 (2013) 074026.
\newblock \href {http://dx.doi.org/10.1103/PhysRevD.87.074026}
  {\path{doi:10.1103/PhysRevD.87.074026}}.

\bibitem{Badger:2005jv}
S.~D. Badger, E.~W. Glover and V.~Khoze, Recursion relations for gauge theory
  amplitudes with massive vector bosons and fermions, JHEP 01 (2006) 066.
\newblock \href {http://dx.doi.org/10.1088/1126-6708/2006/01/066}
  {\path{doi:10.1088/1126-6708/2006/01/066}}.

\bibitem{Alwall:2011uj}
J.~Alwall, M.~Herquet, F.~{Maltoni et al.}, MadGraph 5 : Going beyond, JHEP 06
  (2011) 128.
\newblock \href {http://dx.doi.org/10.1007/JHEP06(2011)128}
  {\path{doi:10.1007/JHEP06(2011)128}}.

\bibitem{Artoisenet:2013puc}
P.~Artoisenet, P.~{de Aquino}, F.~{Demartin et al.}, A framework for Higgs
  Characterisation, JHEP 11 (2013) 043.
\newblock \href {http://dx.doi.org/10.1007/JHEP11(2013)043}
  {\path{doi:10.1007/JHEP11(2013)043}}.

\end{thebibliography}
\end{document}